\documentclass[fleqn,10pt]{wlscirep}
\usepackage[utf8]{inputenc}
\usepackage[T1]{fontenc}
\usepackage{lineno}

\usepackage{graphicx}
\usepackage{multirow}
\usepackage{amsmath}
\usepackage{gensymb}

\DeclareMathOperator*{\argmin}{arg\,min}
\usepackage{subcaption}

\title{A benchmark for 2D foetal brain ultrasound analysis}

\author[1,$\dag$,*]{Mariano Cabezas}
\author[2,$\dag$]{Yago Diez}
\author[3,$\dag$]{Clara Martinez-Diago}
\author[3,$\dag$]{Anna Maroto}
\affil[1]{Brain and Mind Centre, University of Sydney, Sydney, Australia}
\affil[2]{Faculty Of Science, Yamagata University, Yamagata, Japan}
\affil[3]{Hospital Universitari de Girona Doctor Josep Trueta, Girona, Spain}

\affil[*]{corresponding author(s): Mariano Cabezas (mariano.cabezas@sydney.edu.au)}

\affil[$\dag$]{these authors contributed equally to this work}

\begin{abstract}

Brain development involves a sequence of structural changes from early stages of the embryo until several months after birth. Currently, ultrasound is the established technique for screening due to its ability to acquire dynamic images in real-time without radiation and to its cost-efficiency.  However, identifying abnormalities remains challenging due to the difficulty in interpreting foetal brain images. 
In this work we present a set of 104 2D foetal brain ultrasound images acquired during the 20th week of gestation that have been co-registered to a common space from a rough skull segmentation. The images are provided both on the original space and template space centred on the ellipses of all the subjects. Furthermore, the images have been annotated to highlight landmark points from structures of interest to analyse brain development. Both the final atlas template with probabilistic maps and the original images can be used to develop new segmentation techniques, test registration approaches for foetal brain ultrasound, extend our work to longitudinal datasets and to detect anomalies in new images.

\end{abstract}
\begin{document}

\flushbottom
\maketitle

\thispagestyle{empty}


\section*{Background \& Summary}

Foetal brain development is a complex sequence of events ocurrying throughout gestation. From early stages of embryonic development, the brain undergoes structural changes until several months after birth~\cite{Monteagudo,Namburete}. Therefore, understanding normal brain development is essential to identify potential deviations which may lead to neurological disability. Specifically, various studies have described normal milestones within a specific chronology~\cite{Cohen,Kline,Guimaraes}. These changes can be observed by experts in prenatal diagnosis with skills in foetal neurology using ultrasound (US) and magnetic resonance imaging (MRI). However, identifying neurodevelopmental deviations is challenging due to the difficulty in image interpretation~\cite{Guimaraes}.  

Currently, US is the established technique for screening due to its ability to acquire dynamic images in real-time without radiation and cost-efficiently.  The international guidelines recommend the acquisition of a routine first trimester US within the 11th - 14th weeks and a mid-trimester scan within the 19th - 22th weeks of gestation for anatomical evaluation of the foetus~\cite{Salomon}. The standard foetal assessment includes the evaluation of planes acquired using 2D-US. Even though 3D-US is considered useful for prenatal diagnosis of some disorders (mainly involving the face, the skeleton, the cardiovascular system or the brain), the major obstacles for 3D-US implementation worldwide as the main routine acquisition type are related to foetal motion artefacts and acoustic shadowing during volume acquisition. Another widely used prenatal imaging technique is MRI. However, MRI is not used as a primary screening tool and is instead used as a complementary acquisition when foetal abnormalities are suspected (even in selected high-risk cases)~\cite{Griffiths,Prayer}. Moreover, foetal MRI performed before the 18th – 22nd weeks does not usually provide additional information to that obtained by US. Generally, MRI provides a detailed visualization of structures between the 26th and 32nd weeks, being superior to US and less susceptible to limitations from maternal body conditions and foetal presentation (bones do not produce occlusion artefacts in MRI)~\cite{Prayer}. Within this context, a foetal brain US atlas based on 3D scans of healthy foetuses was recently published 
~\cite{Namburete2023}. This atlas demonstrates the feasibility of assessing structural changes in the cortex and in the subcortical grey matter by US. Furthermore, this chronological US-based atlas of the foetal brain at each gestational age has the potential to become a useful tool to detect development abnormalities when combined with the standard planes of mid-trimester routine scans. However, considering that the primary diagnostic tool in pregnancy remains 2D-US, studies aimed at improving the identification of fetal structures should focus on 2D-US.

As there is a degree of variability in the shapes of anatomical regions between individuals, atlases are typically built by taking into account a number of images considered to fall within normal parameters. A crucial step in this process is finding corresponding regions between different images and warping them to a common space, a process known as image registration. In the case of US atlases, the lower signal-to-noise ratio (SNR), the differences in location for the foetuses due to the lack of a robust localization technique and the presence of image artefacts that may blur or shadow salient features and structure boundaries makes this process particularly challenging.

In this paper we present a set of foetal brain 2D-US images with manual landmark annotations of structures of interest and soft probabilistic maps for those structures based on these landmarks in a common space. The aim of this dataset is to provide a set of US images as a starting point to study registration and segmentation of foetal brain scans and provide tools for researchers focusing on foetal brain development. Furthermore, in combination with the recently published 4D atlas, this dataset can be also used to study 2D to 3D US registration and techniques to determine the ``real age'' of gestation and potential abnormalities when comparing to the atlas. To define a common space, a registration algorithm based on fitting an ellipse to the skull segmentation of a convolutional neural network (CNN) was used to align all the images to a common space. In order to avoid misalignements caused by image quality, a subject ellipse was used to estimate an affine transformation to a common (atlas) space. We also took advantage of the automatic differentiation capabilities of the pytorch package in three fundamentally different optimization settings (training a CNN for segmentation, fitting an ellipse to a segmentation boundary, and estimating an affine transformation).

\section*{Methods}






 

In order to create the dataset presented, the data collected was processed as follows: First, the skull was automatically segmented using a UNet network, then an ellipse was automatically fitted to the shape of the skull. These ellipses where used to register each image to a reference image. Soft probabilistic maps where built using the set of registered images. The rest of this section includes details on each of these steps and a summary of other publicly available datasets. A visual summary can be found in Figure~\ref{fig:Methods}.

\subsection*{Ultrasound Data}\label{annotation}

A prospective cohort of low-risk pregnant women was recruited at routine mid-trimester foetal ultrasound scan. All participants initiated antenatal care before the 12th weeks of gestation, underwent the first trimester ultrasound scan between 12th and 14th weeks and had a low risk for aneuploidies in the first trimester combined screening. Written informed consents were obtained from participants. 
A private dataset of 70 pregnant women with a routine mid-trimester foetal ultrasound scan at $\left[20\sim20.6\right]$ weeks without detected abnormalities was acquired, totalling 104 scans (8 women were scanned three times, 18 women were scanned twice and the remaining 44 women were only scanned once). The median of the maternal age was 31 (range 18-42). Images were acquired using high-frequency transabdominal probe (C2-9) of Voluson E10 ultrasound system. For each subject, a transverse view of the foetal head demonstrating a standard normative transcerebellar scanning plane was manually annotated by 2 trained clinical experts (Figure~\ref{fig:Methods} top right). Both experts were part of the Prenatal Diagnosis Unit, one with over 10 years of experience and the other with 5 years of experience. The experts jointly participated in the annotations of each image to highlight common structures related to brain development as follows:

\begin{itemize}
    \item \textbf{Skull}: 4 landmarks from the inner line of the skull were annotated: 2 at the level of the middle line and the other 2 in a perpendicular imagined line at the level of the posterior corners of the \textit{cavum septi pellucidi} (CSP).
    \item \textbf{Thalami}: 1 landmark marking the edges of both thalami at the middle line and the outer edges of the concavity shape were annotated (3 total points).
    \item \textbf{Cerebellum}: 8 landmarks on the perimeter the cerebellum were annotated. Specifically, 2 points from the midline, 2 points from the cerebellum external edges and 4 points in the middle of each cerebellum hemisphere.
    \item \textbf{Cavum septi pellucidi (Cavum)}: 4 points, each marking one corner of its rectangular shape, were annotated.
    \item \textbf{Sylvian Fissure (Sylvius)}: 2 landmarks, one for each sylvian fissure edge, and 1 landmark in the inflection point of the fissure were annotated. For all the images, only the inferior fissure was visible.
    \item \textbf{Midline}: 1 landmark in the upper edge of the midline and 1 landmark in the upper edge of the CSP were annotated.
\end{itemize}

An important aspect of acquisition is that the sylvian fissure was always scanned on the lower part of the image, irrespective of the head orientation (left to right or right to left). This phenomena has important implications for registration. When aligning all images to a common 3D space (for example a 3D atlas template) in order to have all images facing the same direction, the transformation can be modeled with 180$\degree$ rotations over the y axis. If the transformation is limited to the 2D space (image coordinates), a mirroring operation has to be applied to align all images. In the code we provide to process the images, the second option is used for simplicity. Furthermore, our common space is oriented from left to right (anterior to posterior).

Data acquisition was approved by the ethics committee ``Comitè d’Ètica d’Investigació amb Medicaments CEIM GIRONA'' with the code 2023.067. Furthermore, the subjects were informed and consented to the open publication of the data.

\subsection*{Skull segmentation database}
The segmentation dataset used to train a skull segmentation CNN was downloaded from the HC18 grand challenge on ``automated measurement of foetal head circumference using 2D ultrasound images''~\cite{HC18,HC18data}. The challenge comprises a set of $800\times 540$-pixel 2D US images with a pixel sizes $\in(0.052,0.326)$ mm. The data set was split into 999 images for training and 335 for testing. For each image in the training set, an ellipse was manually fitted to the HC by a trained sonographer but precise segmentations of the true skull boundary were not provided. No pre-processing techniques were applied to the images apart from computing the z-score of the intensity values with respect to the mean and standard deviation of the whole image (non-0 intensities) before feeding them to the segmentation network. The network trained with all the training set images was then used to roughly segment the skull of our own 2D US images and provide a starting point to fit an ellipse.

\subsection*{Other public databases}

To the best of our knowledge, only 3 other large datasets focusing on 2-D ultrasound have been made publicly available, including HC18. As mentioned on our brief description of HC18, the annotations are limited (only an ellipse roughly representing the skull is given), the subjects present a large distribution of gestational ages and no further information on the acquired plan e is provided. For studies focusing on the segmentation of structures of interest or registration, new annotations would be needed. Furthermore, the gestational age and acquisition plane for each subject can have a large impact on the appearance of the structures. Another public dataset presented by Burgos-Artizzu et al~\cite{Burgos2020} includes a large set of images (12400) ranging from the 18th to the 40th week of pregnancy. Images were acquired using six different machines and labelled at the image level. Similarly to HC18, the dataset presents a large range of gestational ages and limited annotations that could only be used to develop classification algorithms. Finally, Alzubaidi et al~\cite{Alzubaidi2023} released a public dataset of 3832 high resolution images. In contrast with the other two datasets, no mention of gestational age is provided and once again rough annotations in the form of bounding boxes are provided. Moreover, the authors also highlight image resampling as an additional shortcoming of their dataset.

In our dataset, we focus on a specific acquisition plane and gestational age as defined by international guidelines and raw images are provided. Furthermore, landmarks for the most salient points of each structure of interest are provided with software tools to estimate finer-grained masks and bounding boxes around the landmarks. In that sense, our dataset provides a useful tool to address multiple image analysis problems including registration that could not be easily approached with other available datasets and no additional annotations.

\subsection*{Registration method}

Medical image registration~\cite{Fu,Gholipour} is a necessary initial step for many medical image processing applications that rely on group-wise analysis. Typically, images are registered in pairs: one of them is defined as the ``fixed'' image (or reference) and the other as the ``moving'' image. The moving image is then warped using a transformation function to generate the final ``moved'' image. The transformation is commonly optimised using a predefined metric that computes image similarity between the reference and the moved image after transformation. Here, we present a coarse registration method to roughly align different foetal ultrasound images as described in the following sections and illustrated in Figure~\ref{fig:Methods}. We chose one of the images as the common ``reference'' image (image 10) and registered the remaining images to it.


\subsubsection*{Automatic Skull segmentation using a Unet}\label{elipseSeg}
The Unet architecture~\cite{Ronneberger} is one of the most common CNN architectures for image segmentation. Due to its encoder-decoder structure and the use of skip connections, precise segmentations based on multi-scale features can be obtained for a variety of segmentation applications~\cite{Isensee2021}. Consequently, we used a 2D Unet trained on the HC18 challenge dataset to provide a rough segmentation of the skull for all the images in our registration dataset (see Fig.~\ref{fig:Methods} for an example). Specifically, the encoder and decoder blocks are comprised of 6 convolutional layers with a residual connection~\cite{Kaiming2016} of 32, 32, 128, 256, 256 and 1024 features each (inverse order for the decoder) and a bottleneck of 1024 features. To optimise the weights of the Unet, the Adam algorithm with default initial learning rate was used to minimise the binary cross-entropy loss.

\subsubsection*{Ellipse Registration}\label{elipseReg}

An ellipse is a planar curve representing the locus of the points with constant added distances to two "focal points", as expressed by the  quadratic equation:
\begin{equation}
    f^E(X,Y) = AX^2 + BXY + CY^2 + DX + EY + F = 0, (X, Y)\in \mathcal{R}^2
\end{equation}

\begin{figure}[t]
\begin{tabular}{c}
\subfloat[\textbf{Left}: Automatic Skull segmentation (Unet) example. \textbf{Right}: Ellipse fitted to the segmentation with major, minor axes, its centre and annotated landmarks highlighted.]{\includegraphics[width=\textwidth]{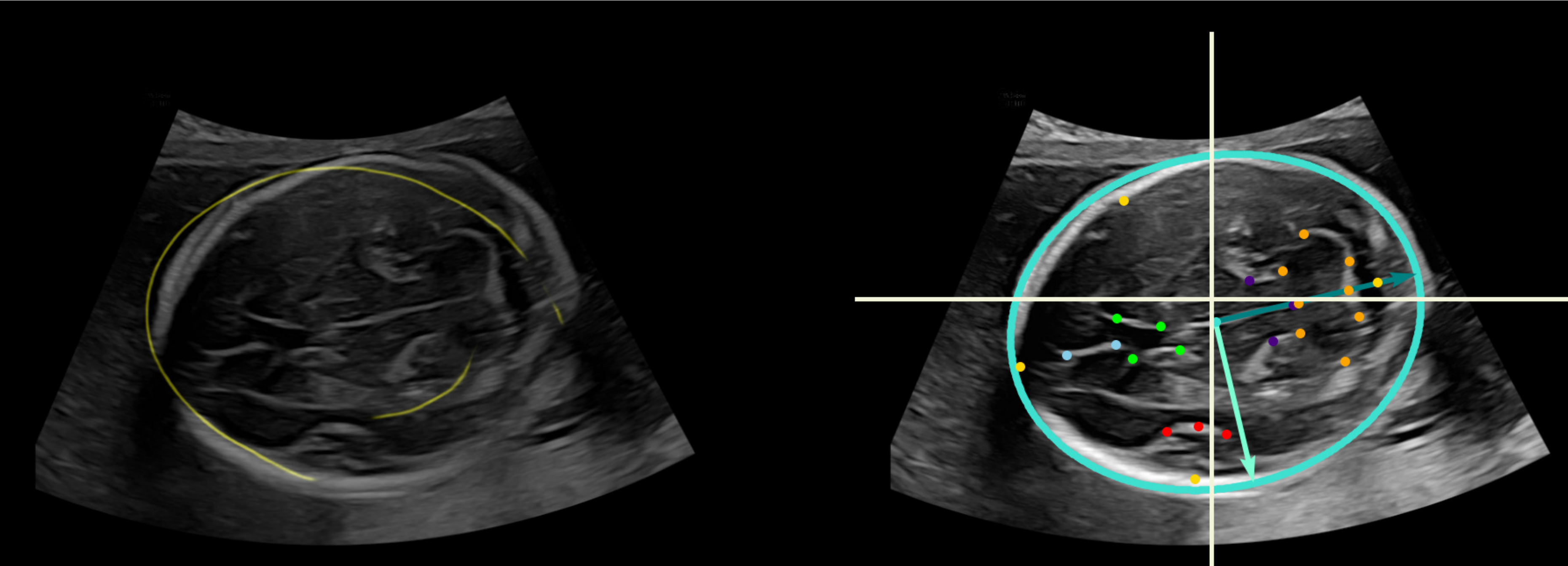}} \\
\subfloat[\textbf{Left}: image transformed after registration. \textbf{Right}: Reference (fixed) image centred.]{\includegraphics[width=\textwidth]{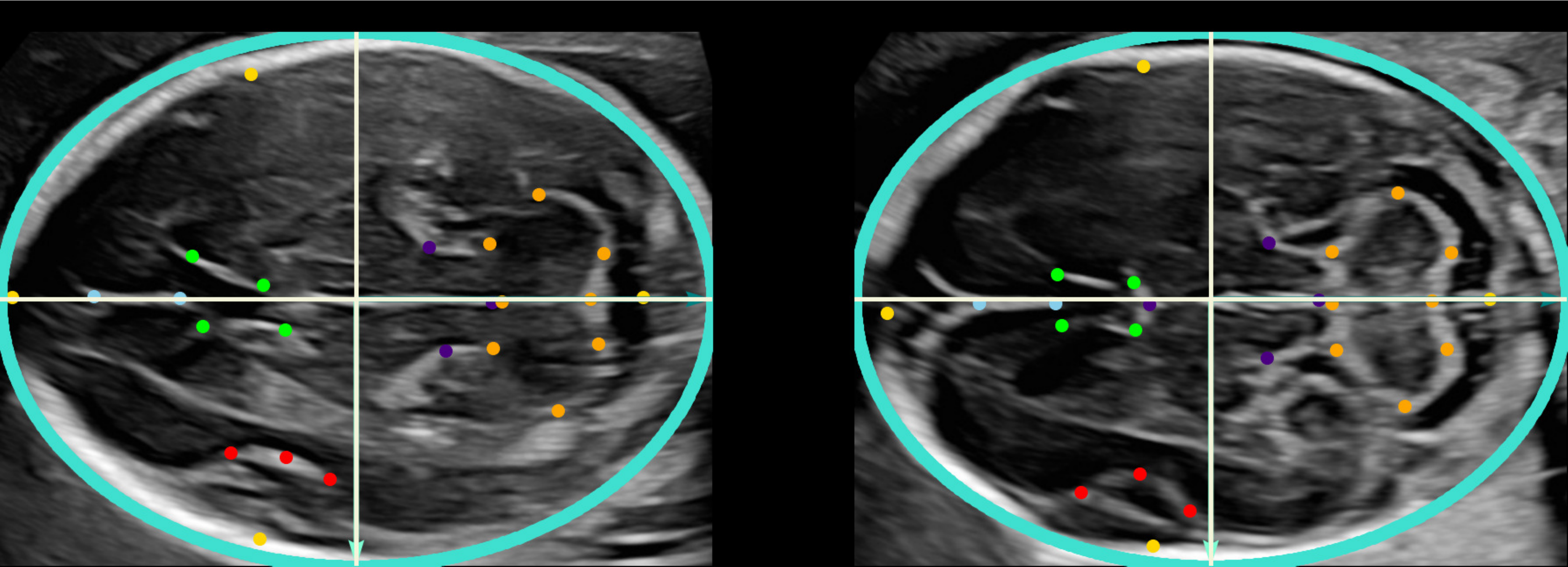}}\\
\end{tabular}
\caption{Methodology: 1) The skull is automatically segmented using a UNet network (\textbf{top-left}). 2) An ellipse is fitted to the skull segmentation (\textbf{top-right}) 3) to estimate an affine transformation to a reference image (\textbf{bottom-right}). The axes of the fitted ellipse are warped to the image coordinate axes and are re-scaled to fit the ellipse in the reference image (\textbf{bottom-left}).}\label{fig:Methods}
\end{figure}

With points that satisfy $f^E(X,Y) \neq 0$ being inside the ellipse perimeter ($f^E(X,Y) < 0$) or outside of it ($f^E(X,Y) > 0$). The general equation's coefficients can be obtained from known semi-major axis $a$ (represented by the magnitude of the turquoise vector in Fig.~\ref{fig:Methods}), semi-minor axis $b$ (represented by the magnitude of the cyan vector in Fig.~\ref{fig:Methods}), centre coordinates $(x_0, y_0)$ (represented by the point where the two vectors meet in Fig.~\ref{fig:Methods}) and rotation angle $\theta$ (the angle from the positive horizontal axis to the ellipse's major axis as observed in Fig.~\ref{fig:Methods}) using the formulae:

\begin{equation}
    A = a^2 \sin^2\theta + b^2 \cos^2 \theta
    \label{ellipse_A}
\end{equation}
\begin{equation}
    B = 2(b^2 - a^2)\sin\theta\cos\theta
    \label{ellipse_B}
\end{equation}
\begin{equation}
    C = a^2\cos^2\theta + b\sin^2\theta
    \label{ellipse_C}
\end{equation}
\begin{equation}
    D = -2Ax_0 - By_0
    \label{ellipse_D}
\end{equation}
\begin{equation}
    E = -Bx_0 - 2Cy_0
    \label{ellipse_E}
\end{equation}
\begin{equation}
    F = Ax^2_0 + Bx_0y_0 + Cy^2 - a^2b^2
    \label{ellipse_F}
\end{equation}

These expressions can be derived from the canonical equation $\frac{x^2}{a^2} + \frac{y^2}{b^2} = 1$ by an affine transformation of the coordinates $(x, y)$ (with a translation $(-x_0, -y_0)$ and an angle $\theta$).
\begin{equation}
    x = (X - x_0)\cos\theta + (Y - y_0)\sin\theta
\end{equation}
\begin{equation}
    y = - (X - x_0)\sin\theta + (Y - y_0)\cos\theta
\end{equation}

Given the set of pixel coordinates of the skull segmentation ($(X_{skull}, Y_{skull}) = \left[(x_1, y_1),\cdots,(x_N, y_N)\right]$) we can fit an ellipse using its parameters ($a$, $b$, $x_0$, $y_0$ and $theta$) by minimising the following objective function:

\begin{equation}
    \hat{a}, \hat{b}, \hat{x_0}, \hat{y_0}, \hat{\theta} = \argmin_{a, b, x_0, y_0, \theta} \sum_{i=1}^N f^E(x_i, y_i)^2,
    \label{eq:objective_function}
\end{equation}

where the coefficients A to F are substituted in $f^E$ by their definitions in equations~\ref{ellipse_A} to~\ref{ellipse_F} and the estimated ellipse parameters.

This process is repeated 5 times, removing erroneous points of the skull segmentation mask that are one standard deviation away from the mean ellipse error according to Eq.~\ref{eq:objective_function}. With the parameters of the skull ellipse estimated through optimisation, we can now define an affine transformation (referred to as \textbf{Ellipse} from now on) to move the brain to the centre of the image as:
\begin{equation}
  \alpha_e = \begin{bmatrix}
    \frac{w}{2a} \cos\theta  & \frac{w}{2a}\sin\theta & - w / 2 \\
    -\frac{h}{2b} \sin\theta  & \frac{h}{2b} \cos\theta & - h / 2 \\
    0  & 0 & 1
\end{bmatrix} \times \begin{bmatrix}
    1  & 0 & - x_0 \\
    0  & 1 & - y_0 \\
    0  & 0 & 1
\end{bmatrix}
\label{eq:ellipse_affine}
\end{equation}






\subsubsection*{Affine image registration}

For comparison, a regular rigid registration of 6 unrestricted parameters was performed with different initialisations. From the most basic identity initialisation (referred to as \textbf{Affine}), to an initialisation using the ellipse parameters of the reference image (referred to as \textbf{Affine (Reference ellipse)}) and a refinement of the ellipse-based affine transformation from Eq.~\ref{eq:ellipse_affine} (referred to as \textbf{Ellipse + Affine}).

\subsubsection*{Probabilistic maps}
Once the images are co-registered to a common space based on their ellipse, a two-step process is performed to generate probabilistic maps for all the structures with more than 2 landmarks.

First, the concave hull of each structure is computed using the alphashapes package. This concave hull represents a rough segmentation of the location of the structure. Second, the segmentations for all the subjects are averaged per structure to determine the probability of each pixel to belong to that structure.

This approach has some limitations. Namely, some structures have a polygonal shape, even though they are actually curves (e.g. sylvius) and the final masks are only a rough representation of the real boundaries  (e.g. cerebellum). However, these polygonal maps can be still used to determine growth milestones and to provide a rough location of the structure of interest.

\section*{Data Records}

The original images with manual landmark annotations (Gimp image editing tool format) and the co-registered images and probabilistic maps used within this paper can be found on figshare~\cite{figshare_data}, and are organised with subject id number (1-52) and an additional number for multiple scans (e.g. 36 and 36.1). Co-registered images are saved in JPEG format with the ``\_registered'' suffix. The final probability maps estimated using a combination of co-registered landmarks and the alphashape package are provided as JPEG images with the name of the structure (e.g. sylvius.jpeg), while comma separated value (CSV) files with the point landmarks of all registered subjects are compressed into a single zip file.

In total, the released dataset consists of 104 annotated 2D US images of foetal brains on the 20th week of pregnancy. The manual annotations are described in \textbf{Ultrasound dataset} section.

\section*{Technical Validation}


To validate the techniques used for co-registration to a common space, we focused on common medical imaging metrics for registration using landmarks. In this section we describe these metrics and provide some qualitative and quantitative results of the alignment (including a visualisation of the probabilistic maps of one of the structures of interest).

\begin{figure}[h]
    \begin{subfigure}{0.5\textwidth}
        \centering
        \includegraphics[width=0.9\linewidth]{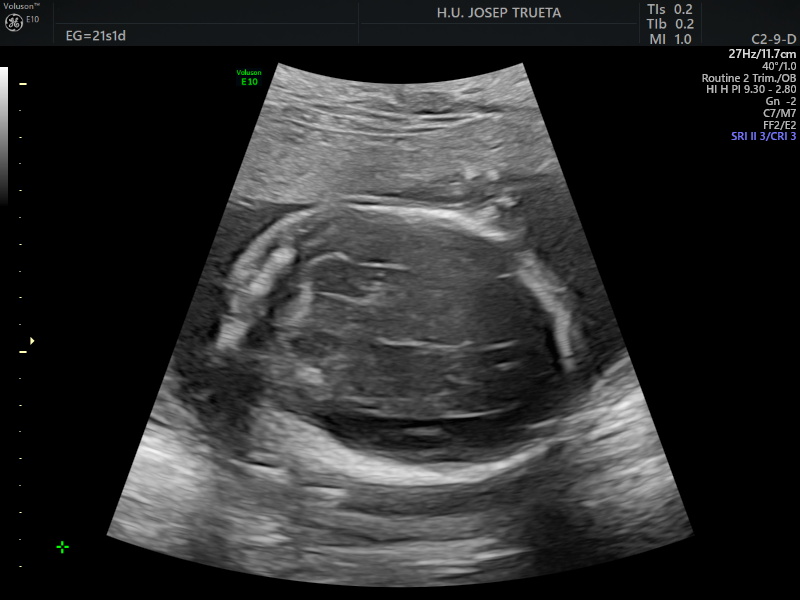}
        \caption{Lower quality image}
        \label{fig:primer}
    \end{subfigure}
     \begin{subfigure}{0.5\textwidth}
        \centering
        \includegraphics[width=0.9\linewidth]{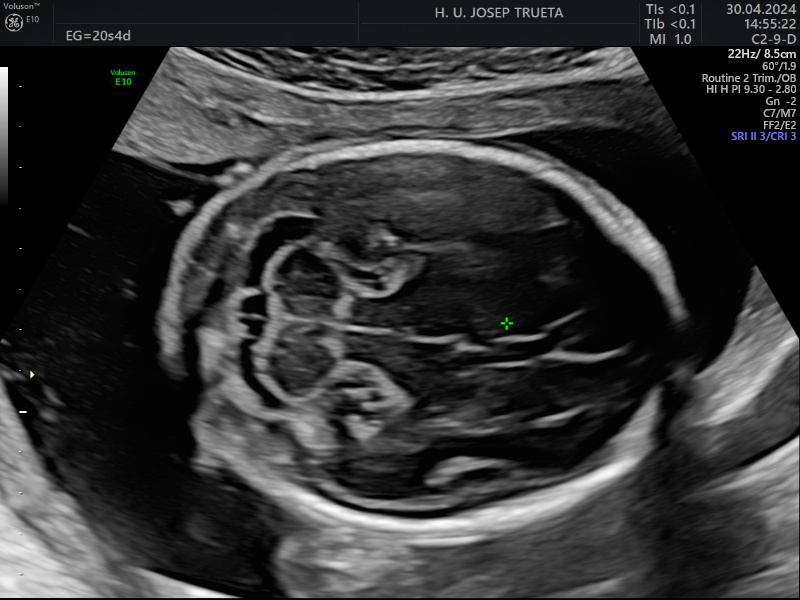}
        \caption{Image from the dataset}
        \label{fig:image_quality}
    \end{subfigure}  
    \caption{Qualitative comparison between a low quality image where structures are not clearly visible and and image from the dataset.}
\end{figure}

Regarding the quality of the US images, all images were acquired with a high frequency ultrasound probe (2-9 MHz) that provides high resolution images following the ISUOG recommendations. The most constraining factor for the quality of the ultrasound images is maternal obesity. However, for this study we excluded women with maternal morbid obesity (body mass index > 40) as it is one of the factors for high-risk pregnancy. To further illustrate that point, we provide a comparative example between an image from the dataset and a lower quality one in figure~\ref{fig:image_quality}.

\subsection*{Evaluation metrics}

For this study we chose to use the anatomical knowledge defined by expert annotations as the main directive to evaluate the quality of the registration and avoid focusing exclusively on common pixel-metrics that might be unreliable and disconnected from physical properties~\cite{Rohlfing}. Specifically, we used the point annotations described in the \textbf{Ultrasound Data} section with two point-based metrics and one area-based metric for every anatomical structure with more than 2 points (all the structures except the midline). For completeness sake, we also included the structural similarity index metric (SSIM) pixel-based metric.

\noindent {\bf Point-based metrics}
In order to penalise partial matches between anatomical structures, we considered the {\it Hausdorff distance} ($d_H$), that computes the worst possible Euclidean distance ($d(p_i, p_j)$) between two sets of points $P_f$ ($|P_f| = N$) and $P_m$ ($|P_m| = M$). We also considered the {\it average of the minimum Euclidean distances} ($d_E$) to express global similarity between structures defined by landmarks.




\noindent {\bf Area-based metric}
To evaluate the superposition between two anatomical structures defined as 2D landmarks (points), we first computed their concave hulls and then considered their Dice similarity coefficient (polygon DSC).



\noindent {\bf Image similarity metric}
For completeness, we also considered the structural similarity index measure (SSIM) as a pixel-intensity-based metric.

\subsection*{Comparison of coarse registration approaches}

\begin{figure}[t]
\begin{tabular}{cc}
\subfloat[Skull]{\includegraphics[width=0.45\textwidth]{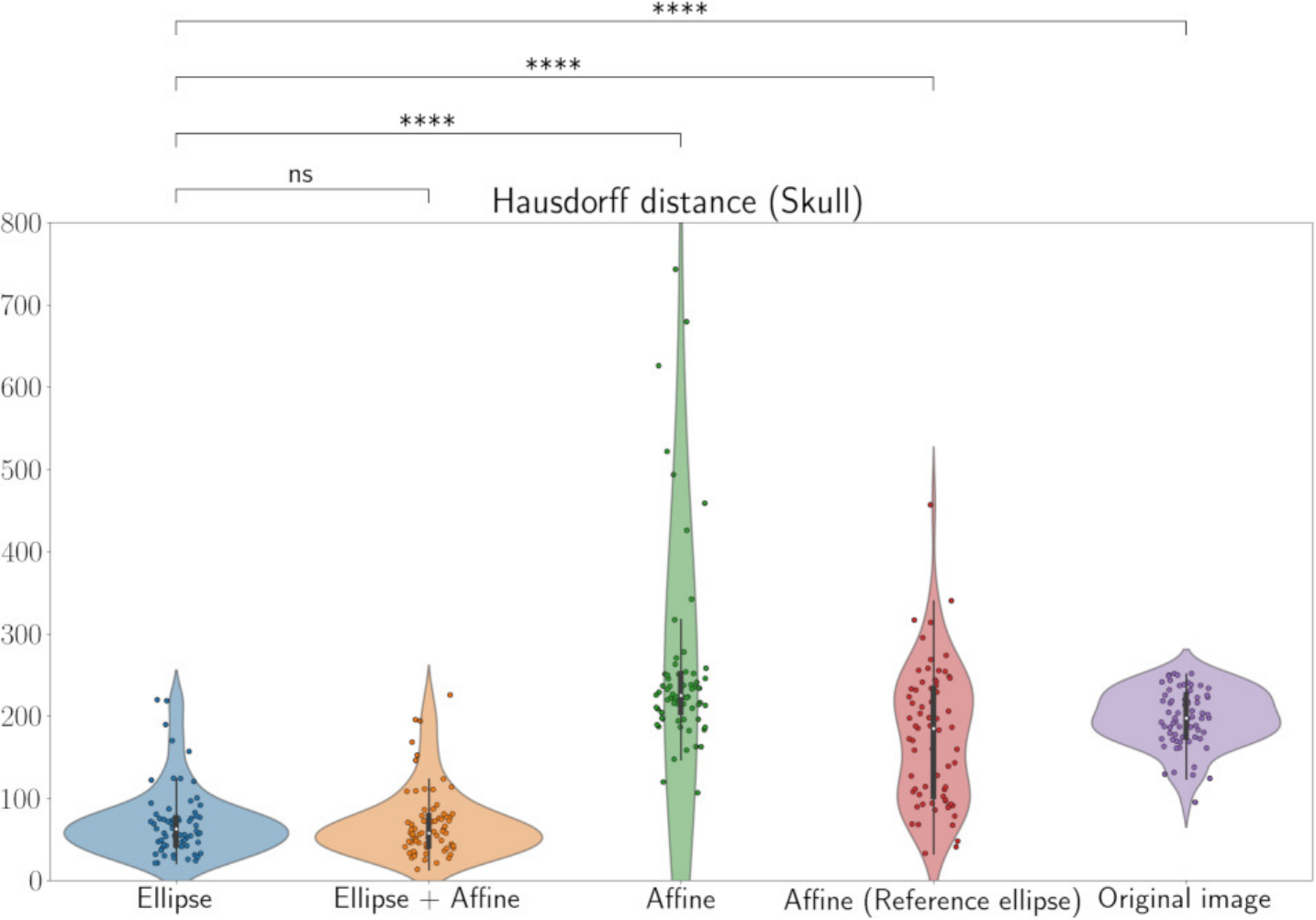}} & \subfloat[Thalami]{\includegraphics[width=0.45\textwidth]{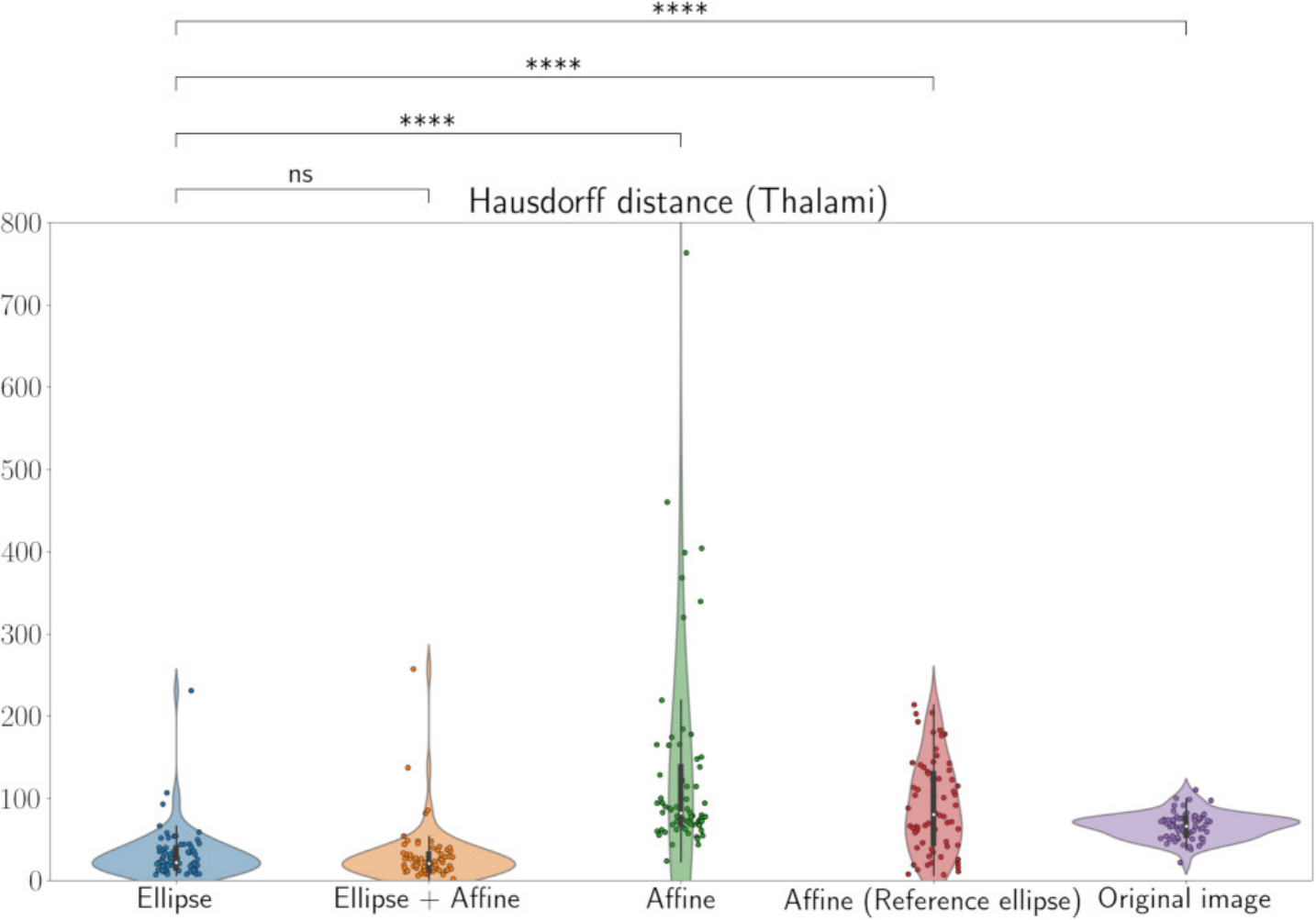}} \\
\subfloat[Cerebellum]{\includegraphics[width=0.45\textwidth]{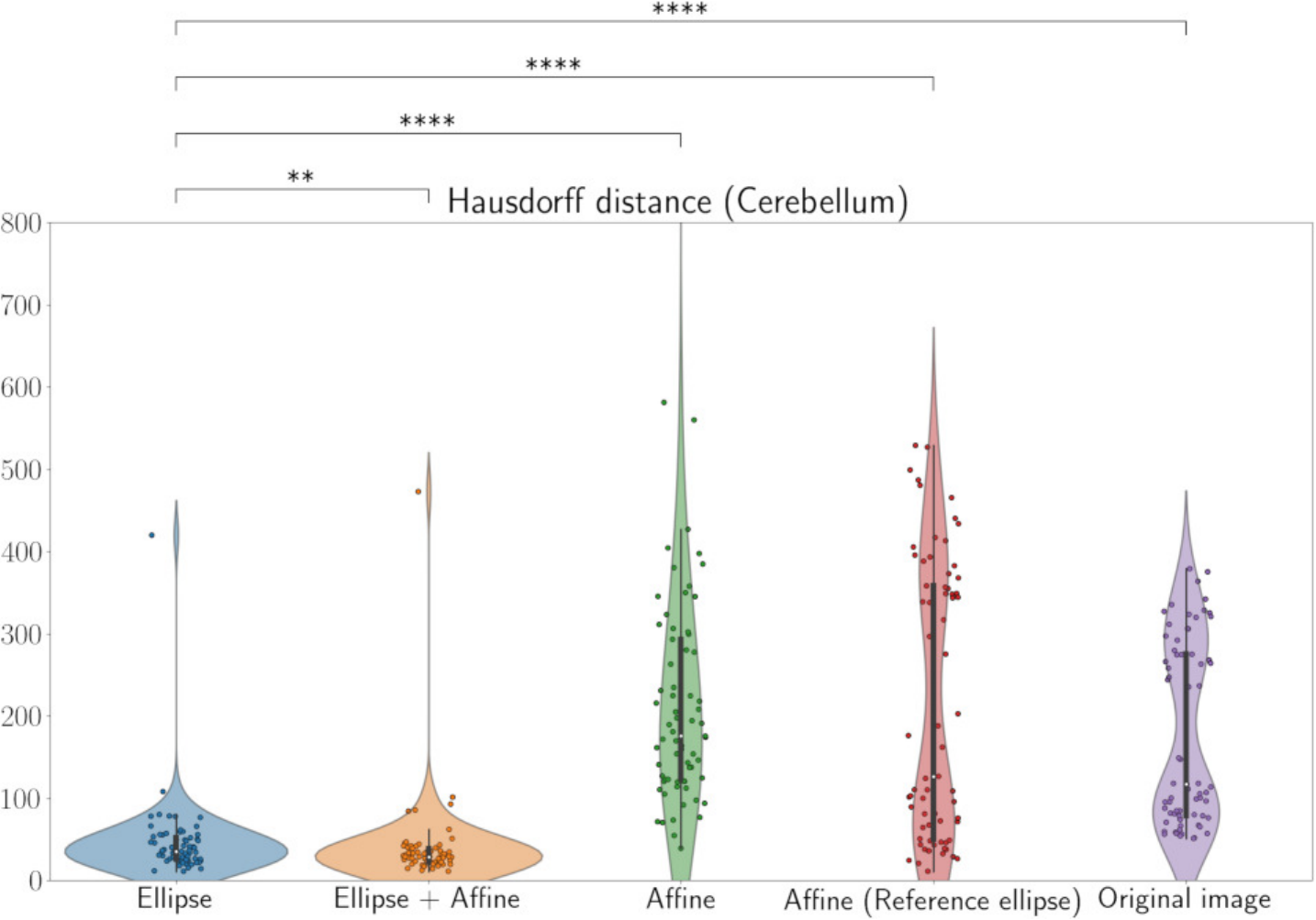}} & \subfloat[Cavum]{\includegraphics[width=0.45\textwidth]{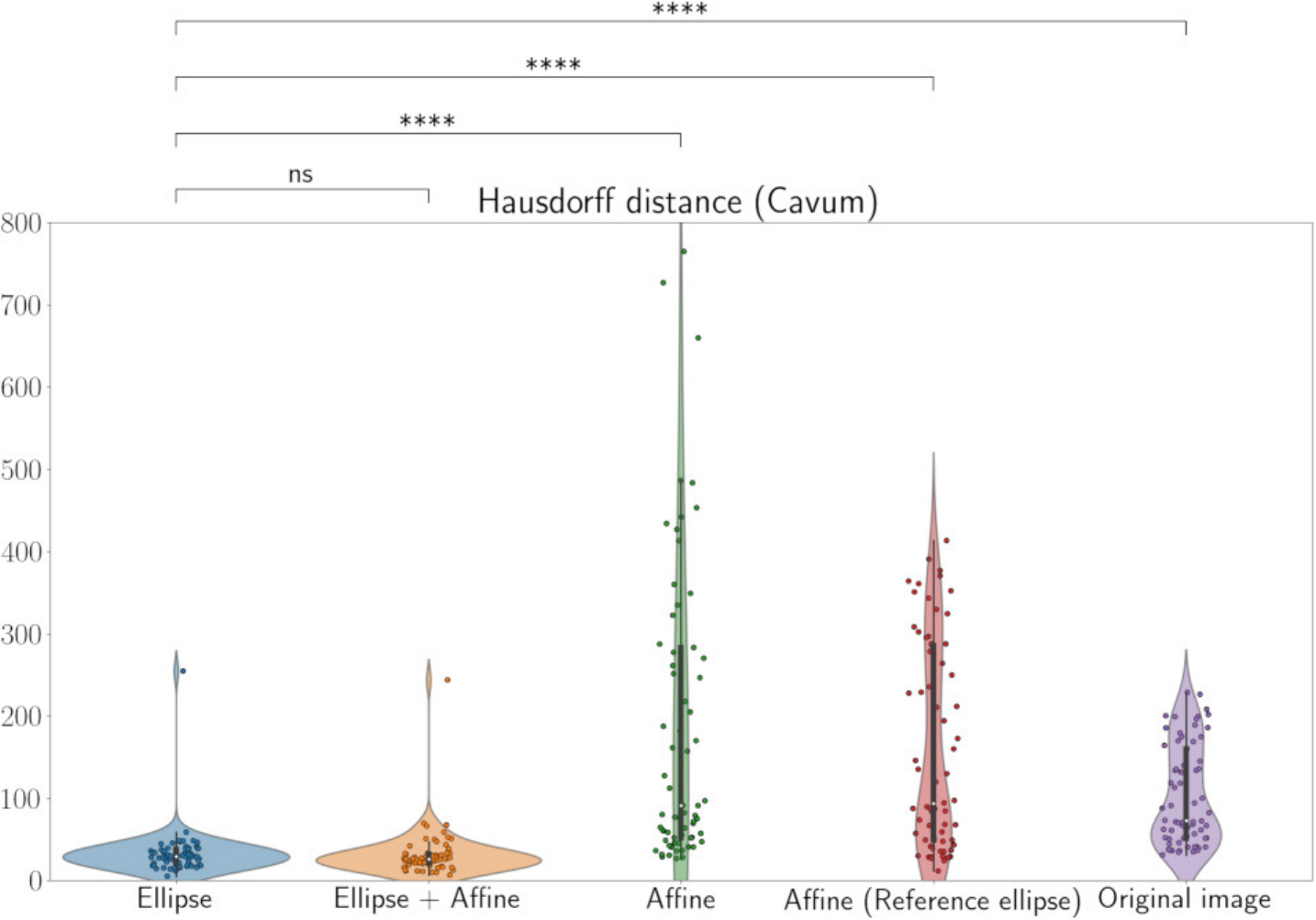}} \\
\subfloat[Sylvius]{\includegraphics[width=0.45\textwidth]{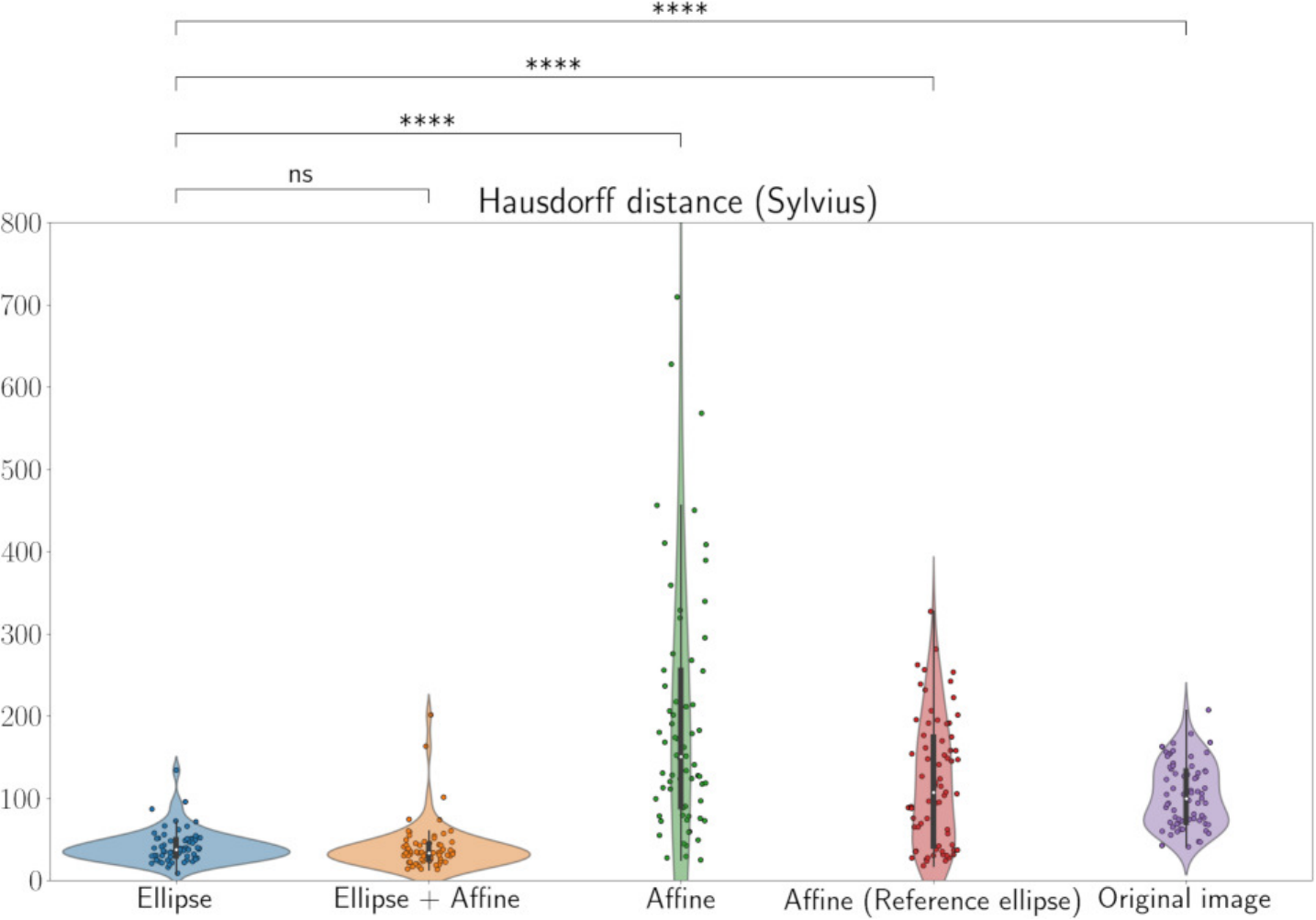}} & \subfloat[Midline]{\includegraphics[width=0.45\textwidth]{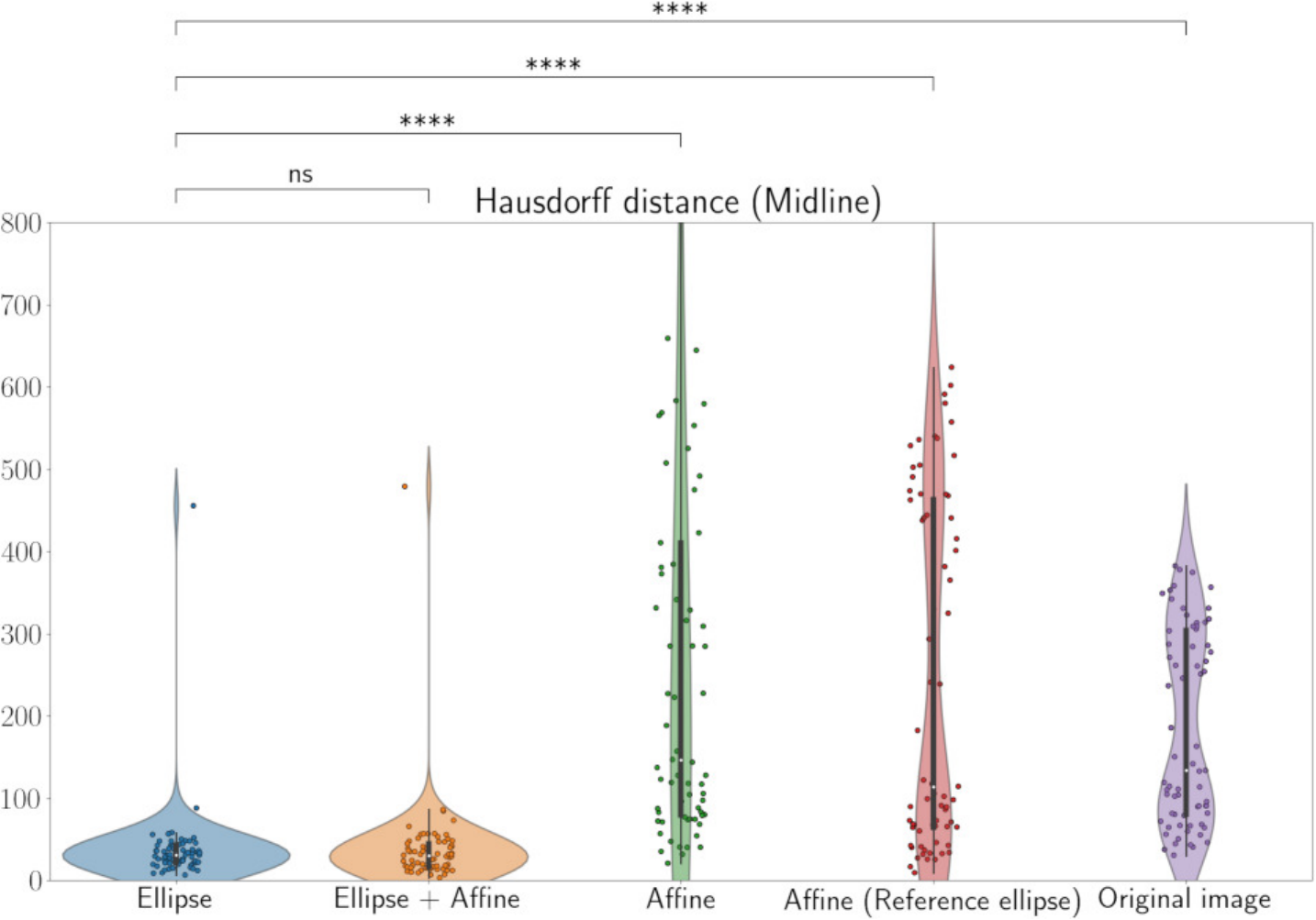}} \\
\end{tabular}
\caption{Quantitative results for all the methods on each structure (Hausdorff distance, $d_H$, lower values indicate better registration). The upper part of each boxplot figure indicates the results of pairwise statistical Wilcoxon tests: (ns: 5.00e-02 $<$ p $\leq$ 1.00e+00, *: 1.00e-02 $<$ p $\leq$ 5.00e-02, **: 1.00e-03 $<$ p $\leq$ 1.00e-02, ***: 1.00e-04 $<$ p $\leq$ 1.00e-03, ****: p $\leq$ 1.00e-04).}\label{fig:results_hausdorff}
\end{figure}

\begin{figure}[t]
\begin{tabular}{cc}
\subfloat[Skull]{\includegraphics[width=0.45\textwidth]{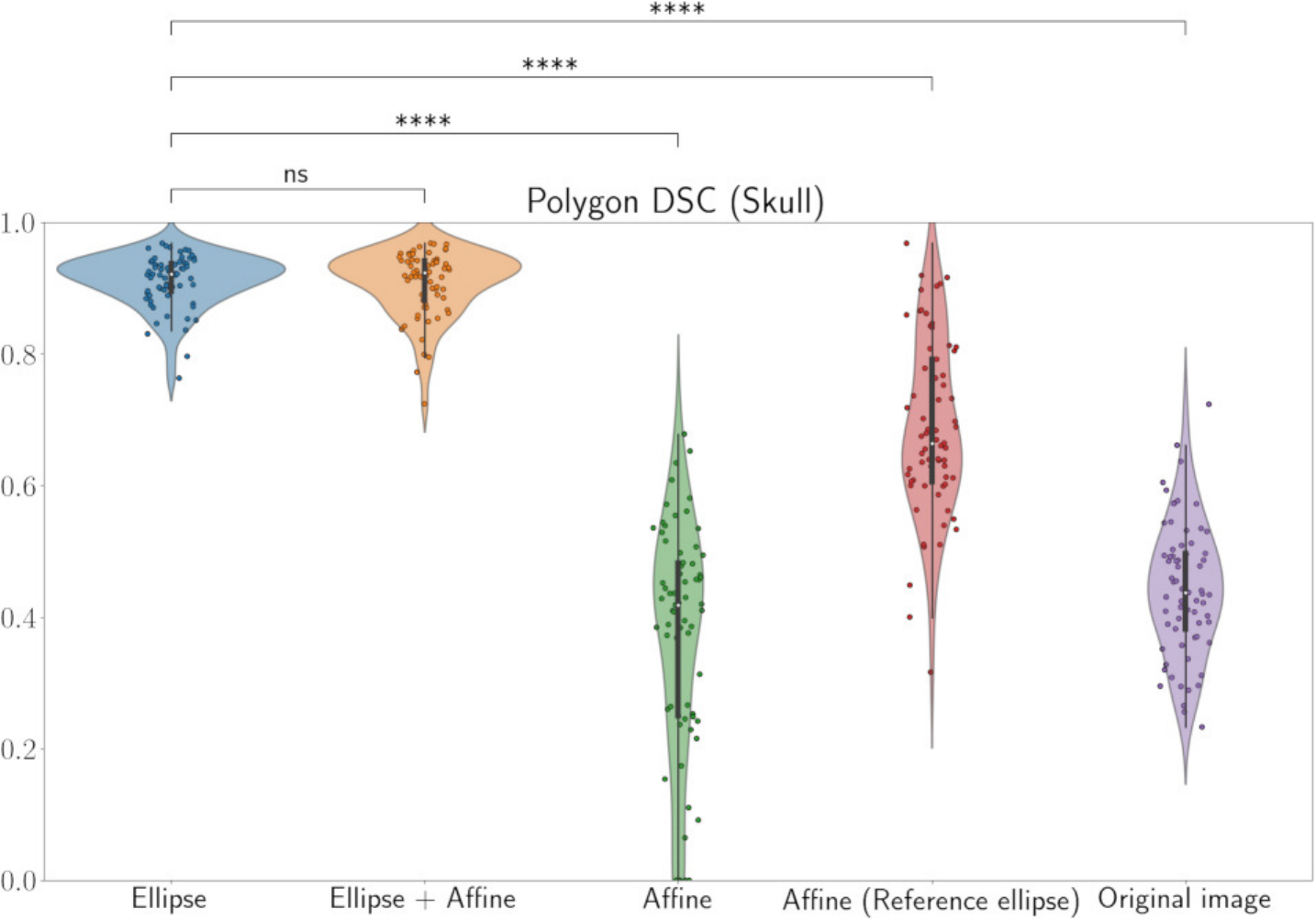}} & \subfloat[Thalami]{\includegraphics[width=0.45\textwidth]{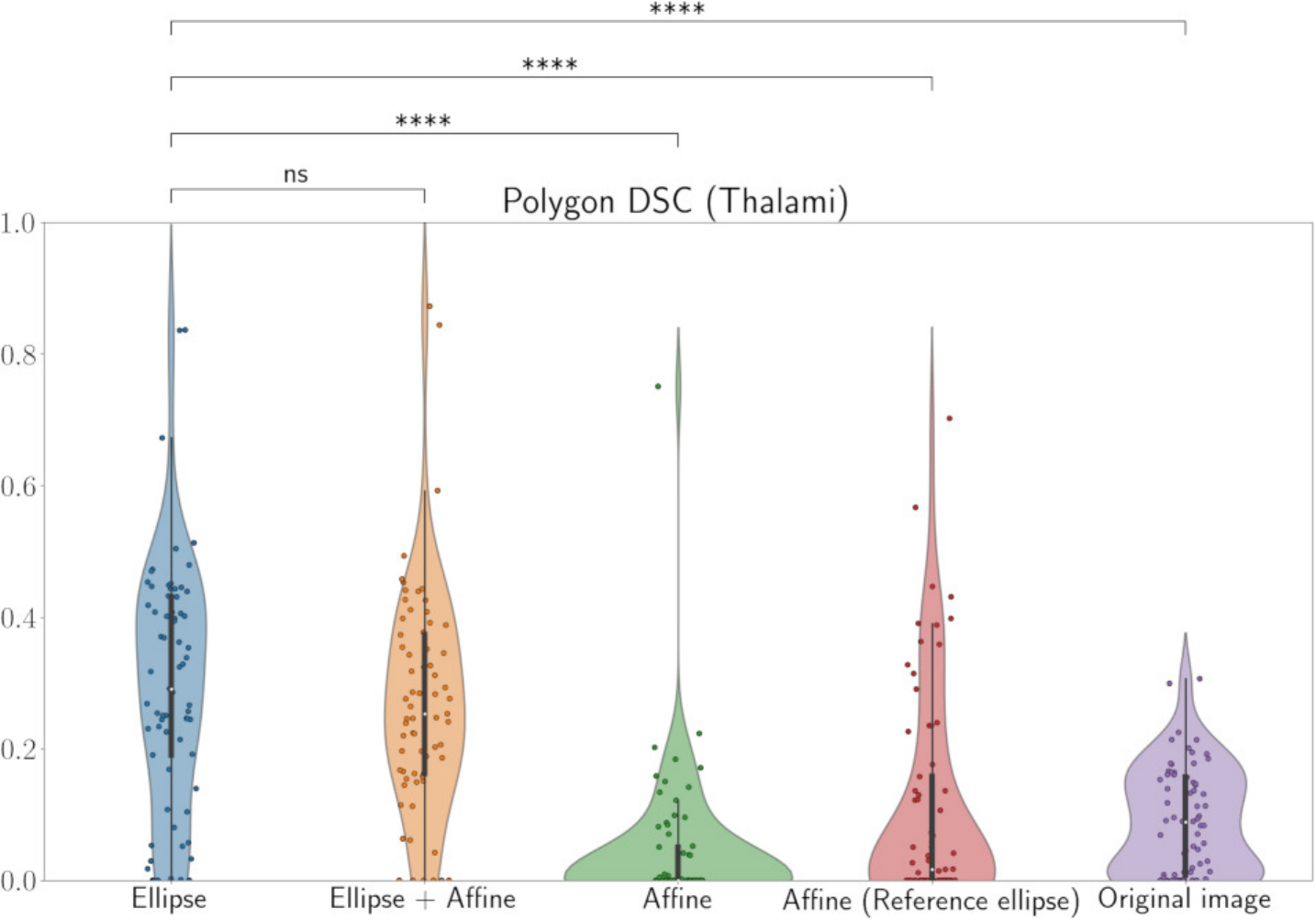}} \\
\subfloat[Cerebellum]{\includegraphics[width=0.45\textwidth]{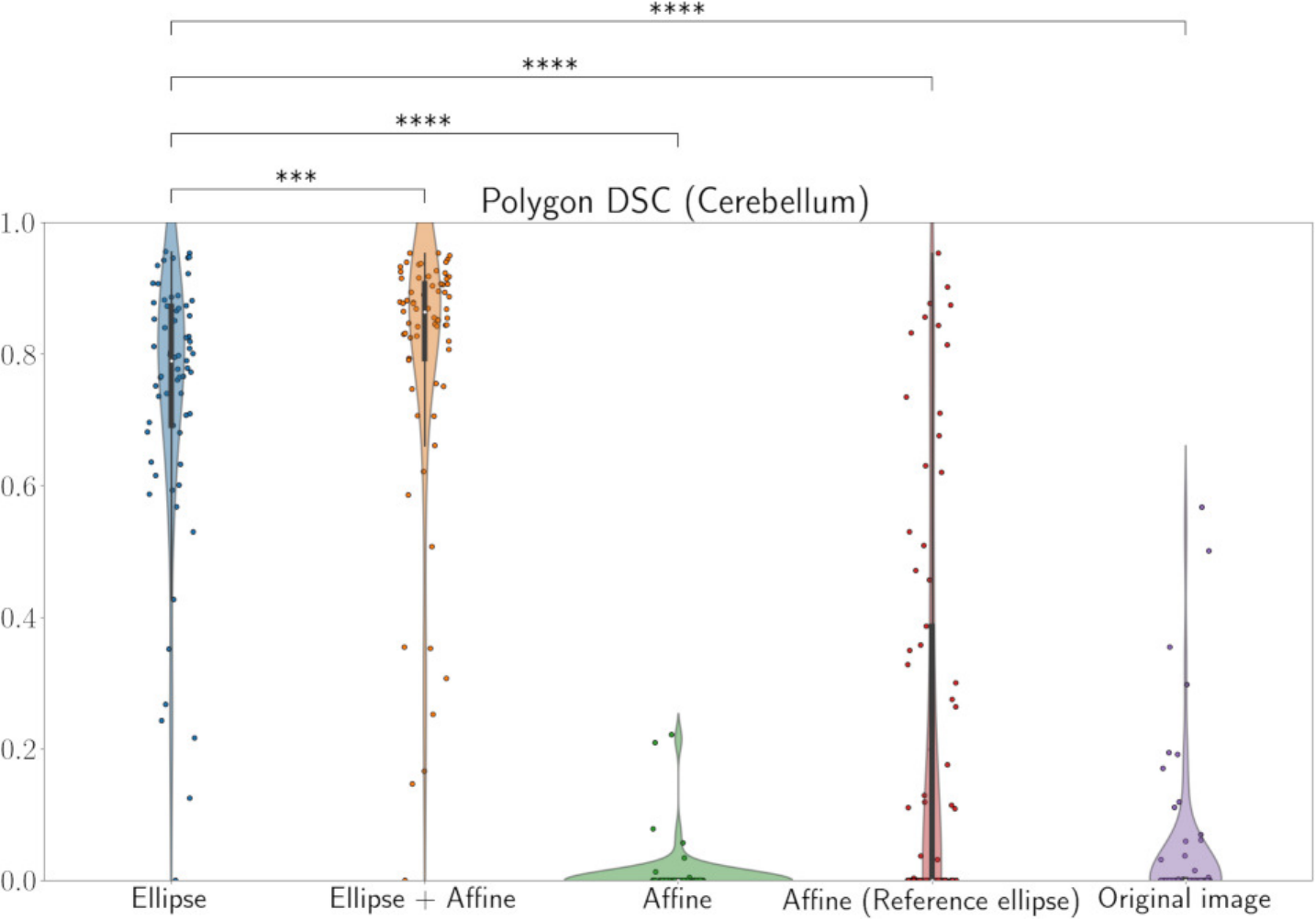}} & \subfloat[Cavum]{\includegraphics[width=0.45\textwidth]{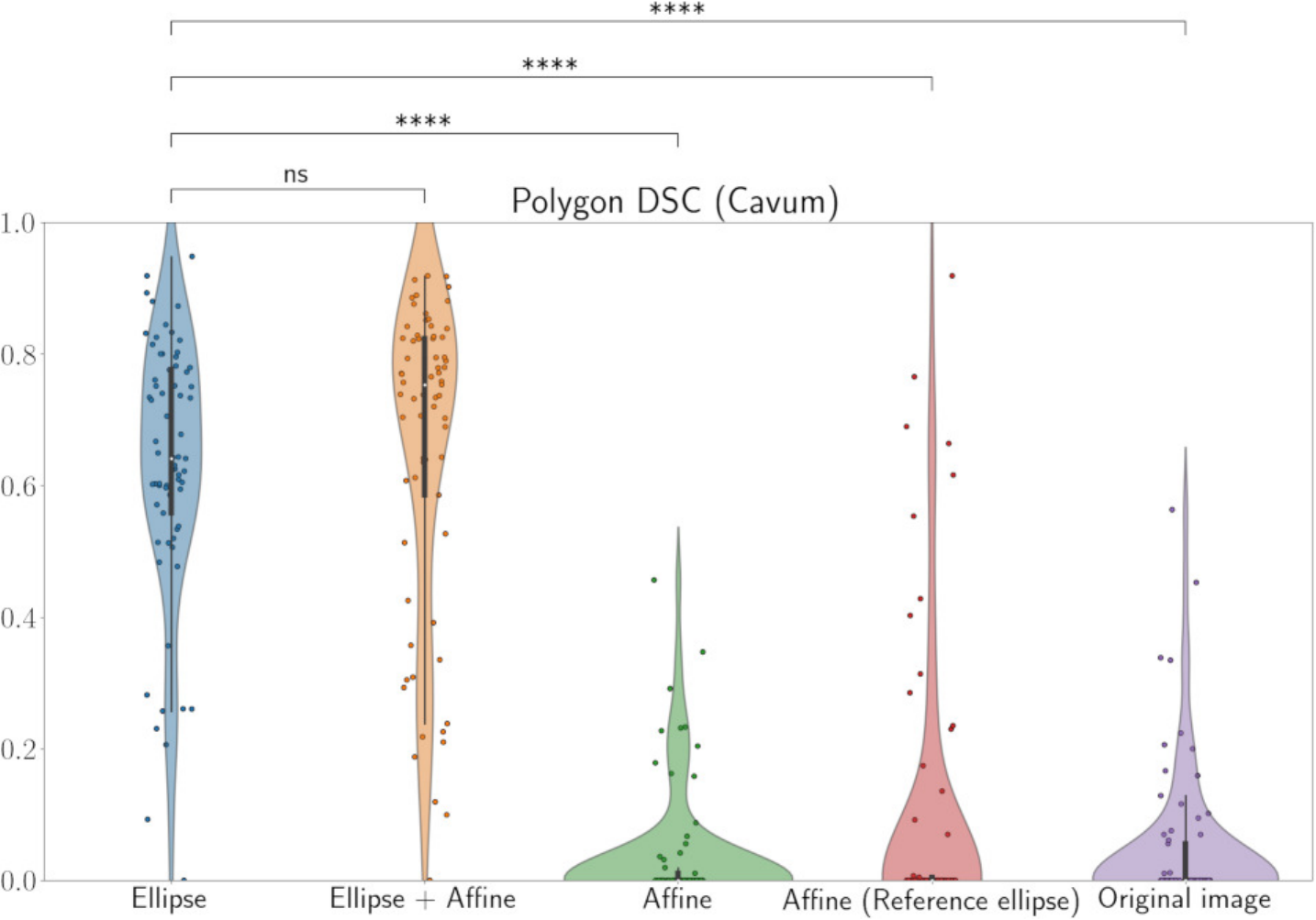}} \\
\end{tabular}
\caption{Quantitative results for all the methods on each structure with multiple points (polygon DSC, higher values indicate better registration). The upper part of each boxplot figure indicates the results of pairwise statistical Wilcoxon tests: (ns: 5.00e-02 $<$ p $\leq$ 1.00e+00, *: 1.00e-02 $<$ p $\leq$ 5.00e-02, **: 1.00e-03 $<$ p $\leq$ 1.00e-02, ***: 1.00e-04 $<$ p $\leq$ 1.00e-03, ****: p $\leq$ 1.00e-04).}\label{fig:results_pdsc}
\end{figure}

\begin{figure}[t]
\centering
\includegraphics[width=0.45\textwidth]{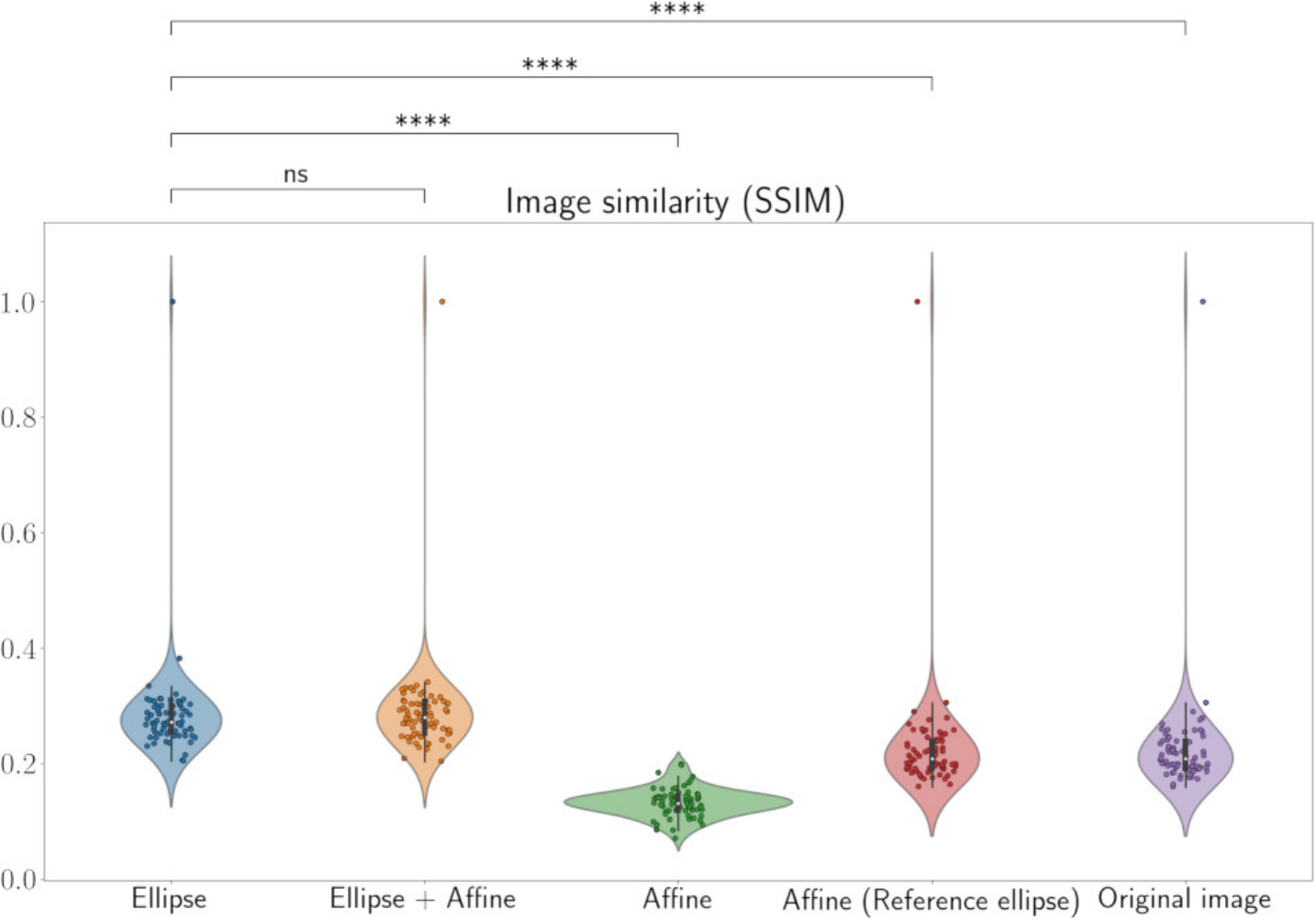}
\caption{Quantitative results for all the methods according to the SSIM metric (higher values indicate better registration). The upper part of each boxplot figure indicates the results of pairwise statistical Wilcoxon tests: (ns: 5.00e-02 $<$ p $\leq$ 1.00e+00, *: 1.00e-02 $<$ p $\leq$ 5.00e-02, **: 1.00e-03 $<$ p $\leq$ 1.00e-02, ***: 1.00e-04 $<$ p $\leq$ 1.00e-03, ****: p $\leq$ 1.00e-04).}\label{fig:results_ssim}
\end{figure}

\begin{figure}[t]
\begin{tabular}{cc}
\subfloat[Ellipse registration]{\includegraphics[width=0.45\textwidth]{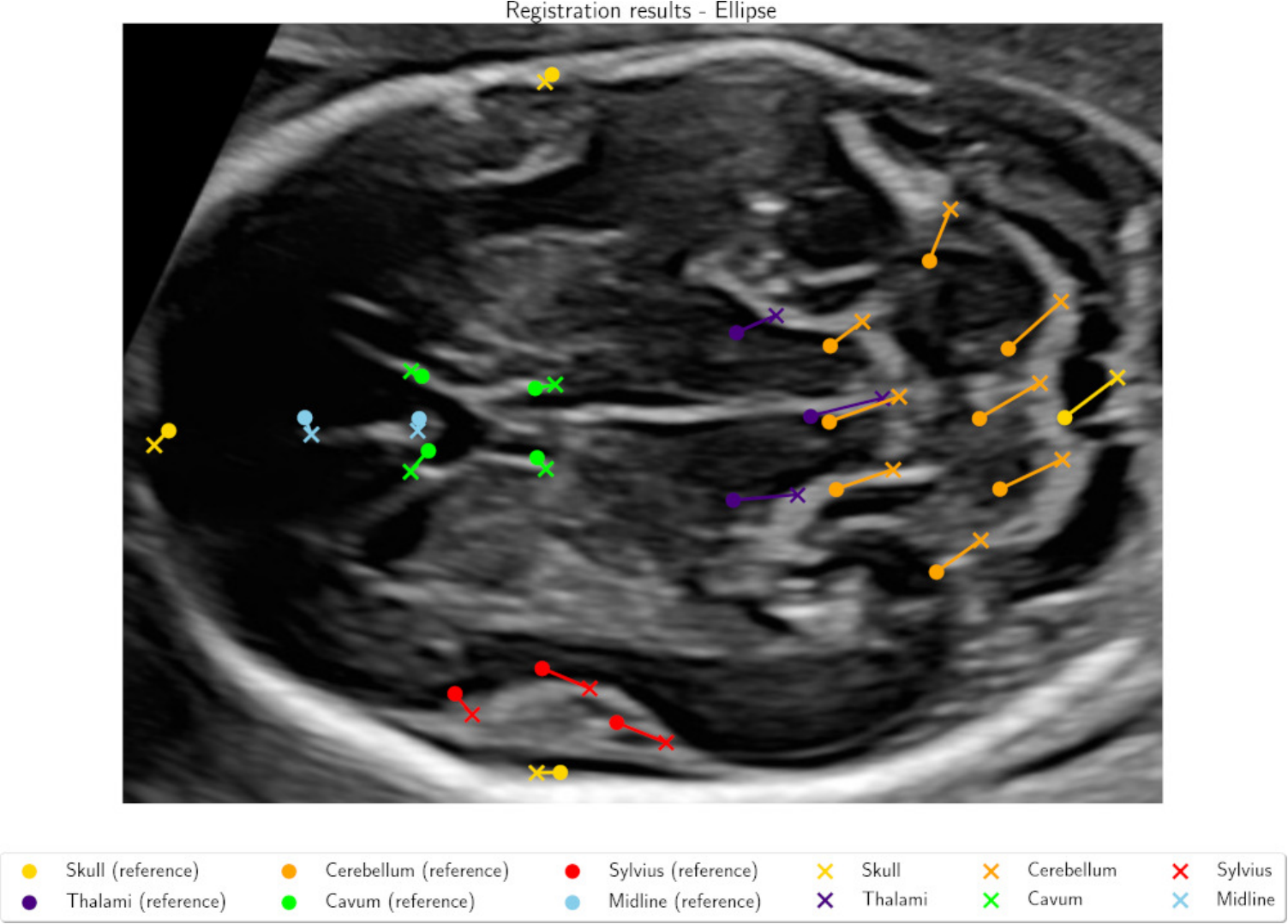}} & \subfloat[Ellipse + Affine registration]{\includegraphics[width=0.45\textwidth]{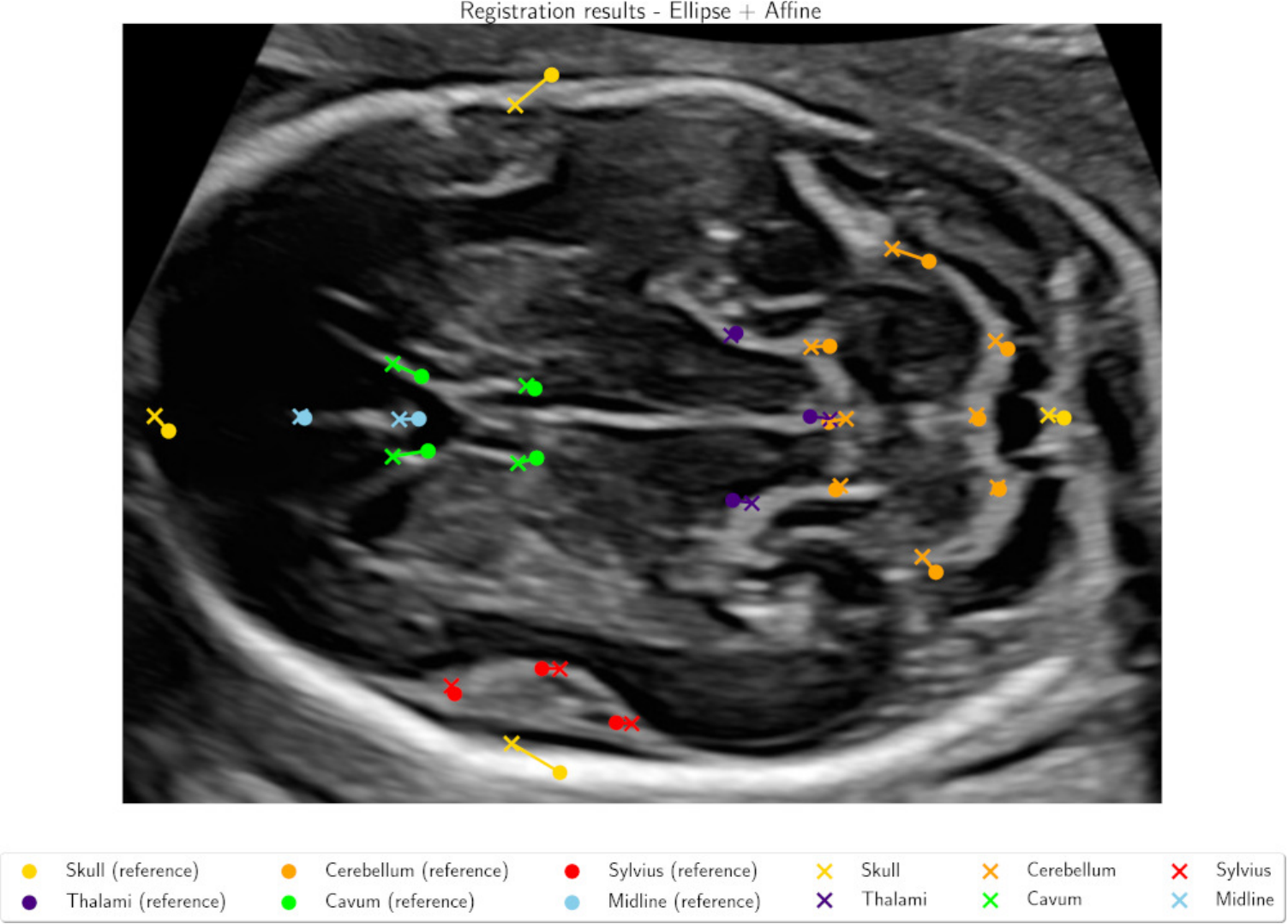}} \\
\subfloat[Affine registration]{\includegraphics[width=0.45\textwidth]{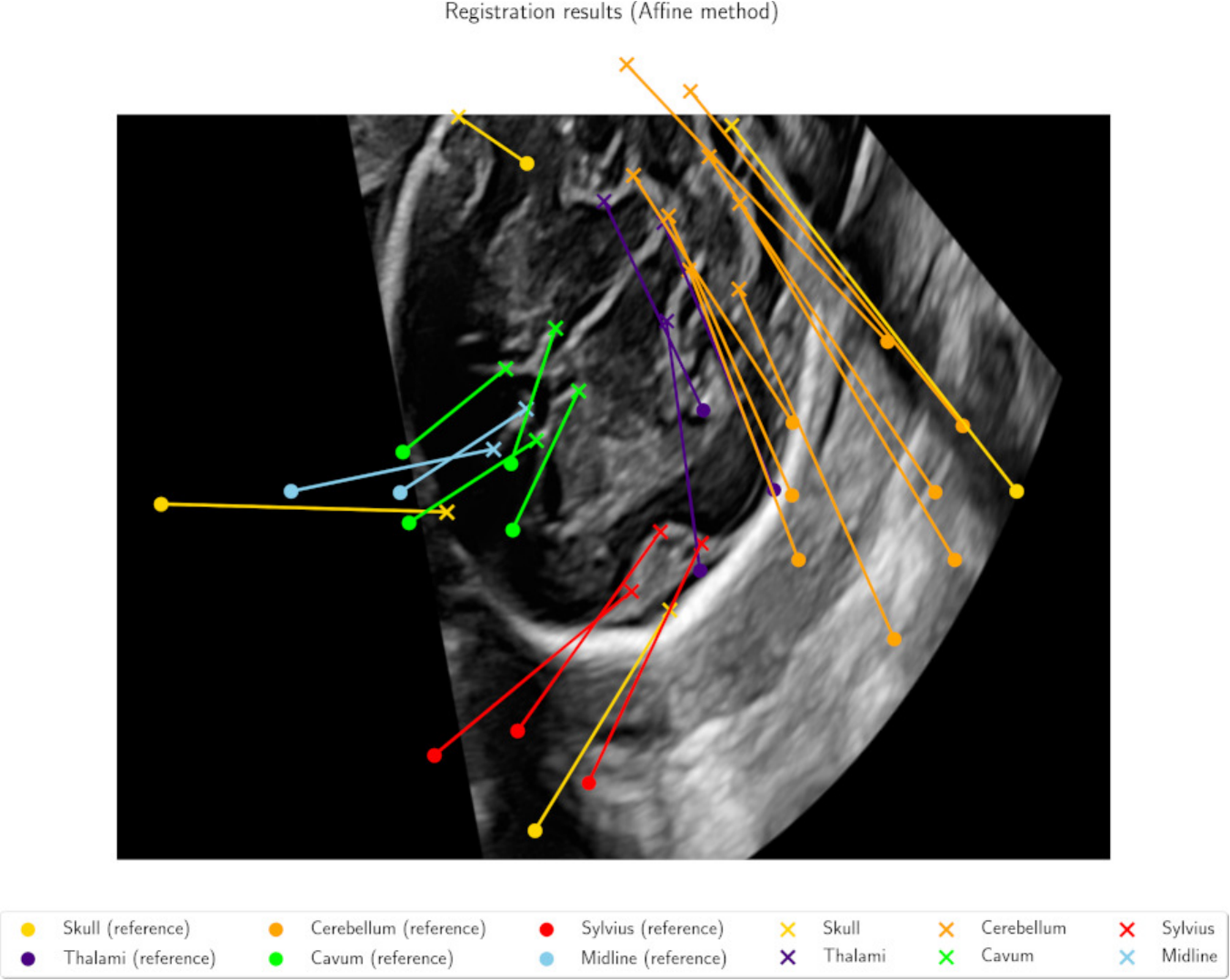}} & \subfloat[Affine (reference ellipse) registration]{\includegraphics[width=0.45\textwidth]{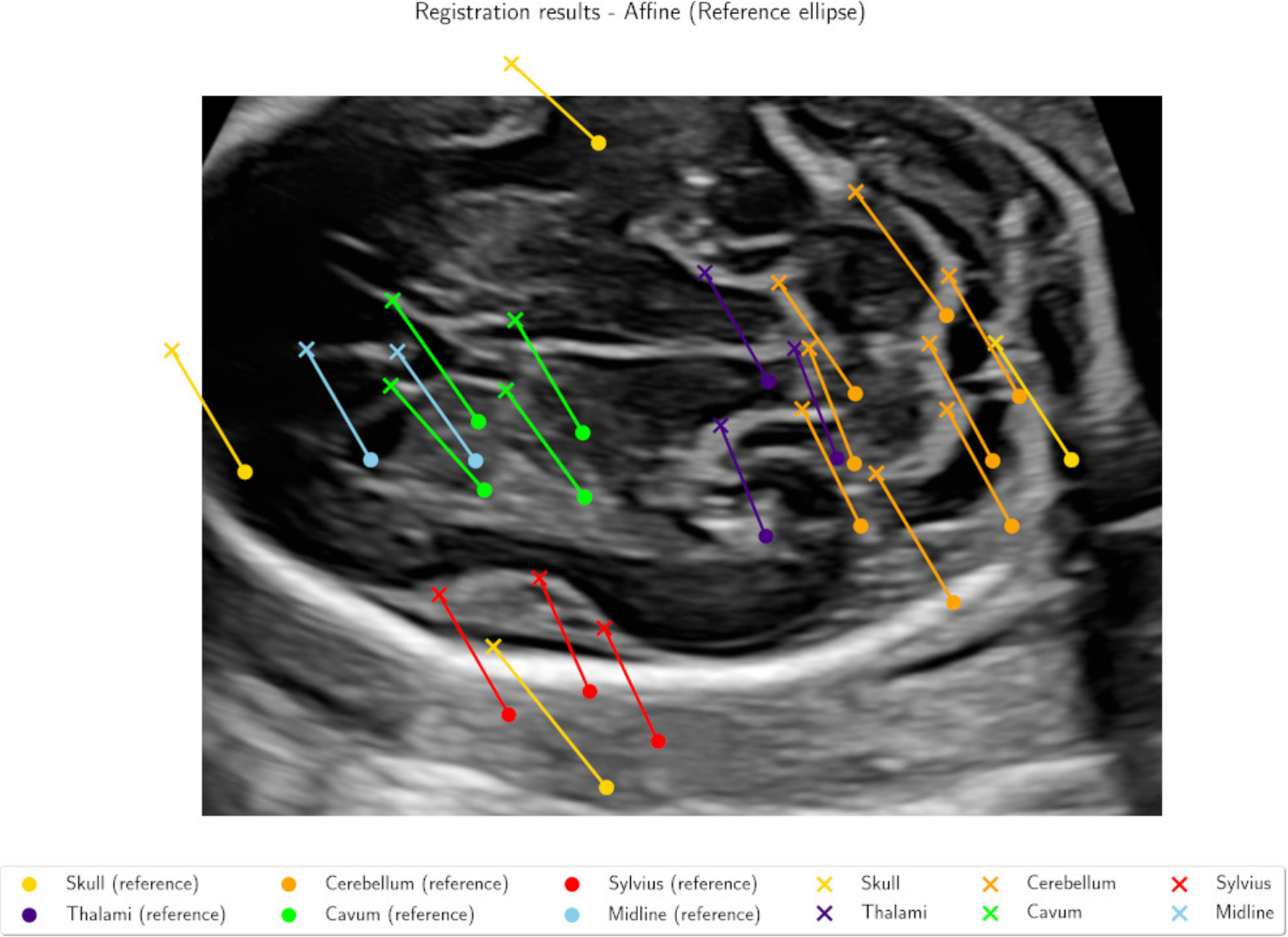}} \\
\subfloat[Original image]{\includegraphics[width=0.45\textwidth]{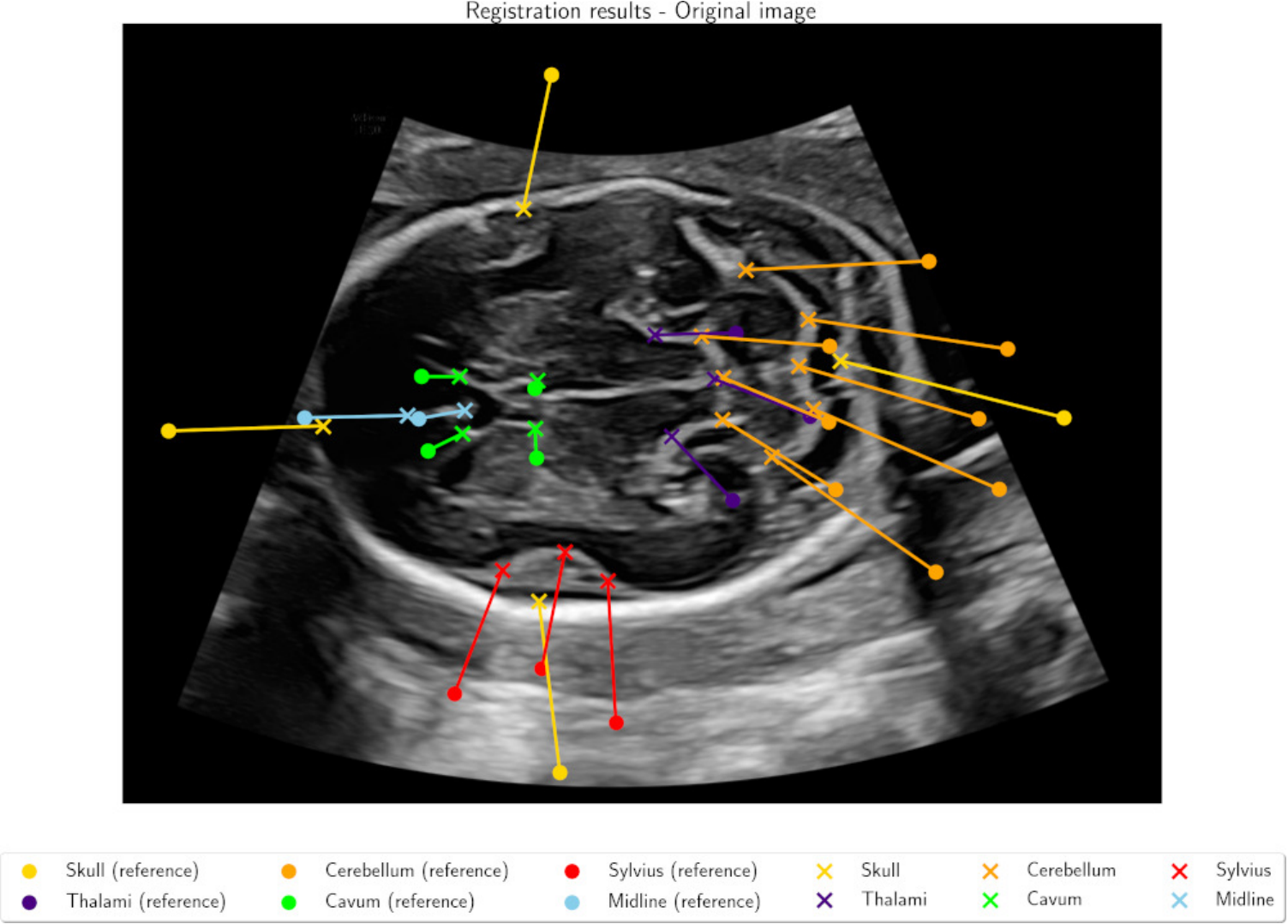}} & \subfloat[Reference image]{\includegraphics[width=0.45\textwidth]{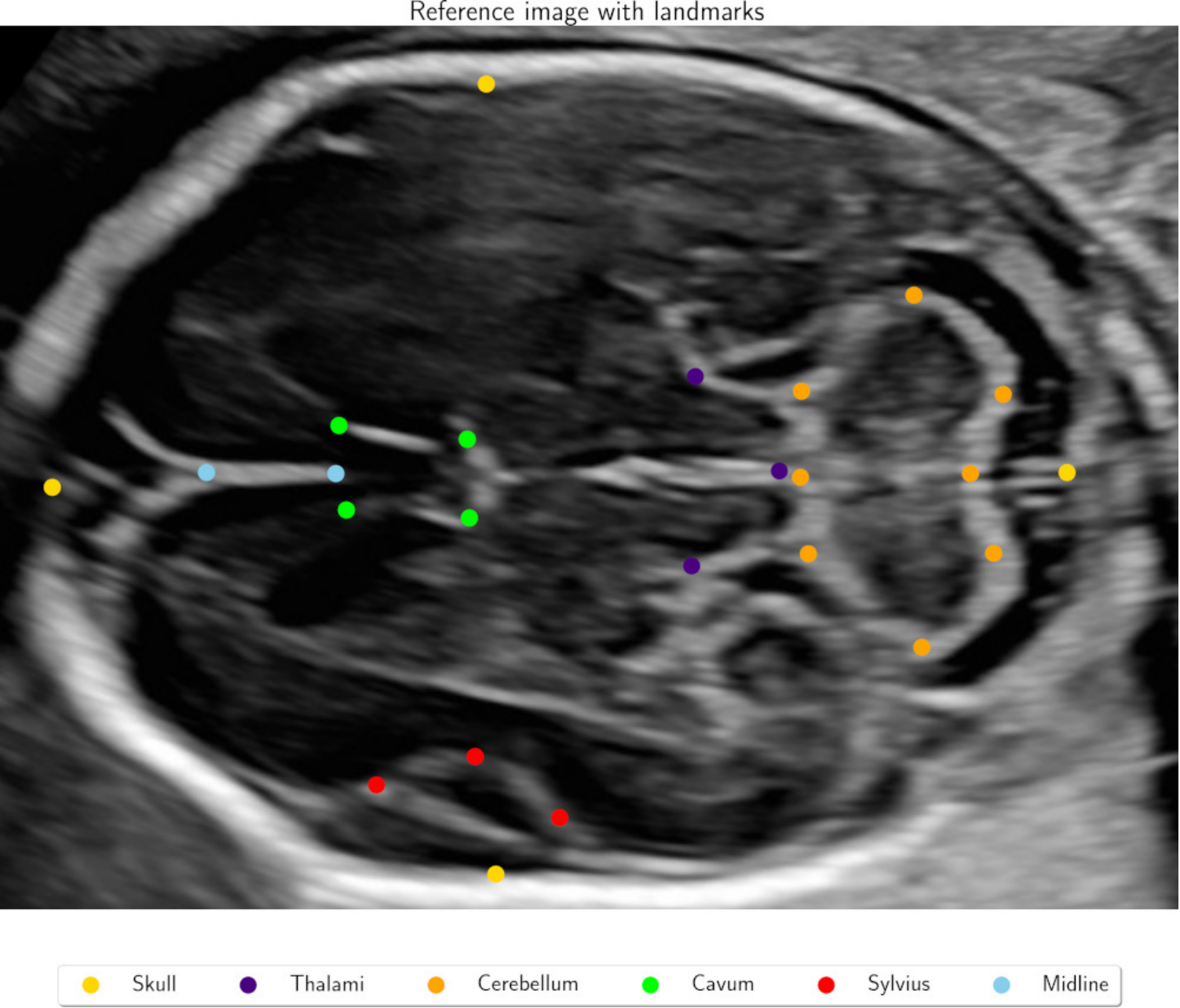}} \\
\end{tabular}
\caption{Qualitative example of the alignment between the reference image and a randomly selected subject (11). The background image corresponds to the warped image after registration, circles denote the reference points for each structure, crosses represent the registered points and lines are used to illustrate misalignment between the reference and the registration.}\label{fig:reg_scatter}
\end{figure}

Figures~\ref{fig:results_hausdorff},~\ref{fig:results_pdsc} and~\ref{fig:results_ssim} summarise the results with boxplots and Wilcoxon signed-rank tests for all the registration methods and metrics considered. Wherever possible, the results for different anatomical areas are presented separately. For the Euclidean and Hausdorff metrics, lower values indicate better registration, while for the SSIM and Dice metric higher results indicate better registration. Moreover, a qualitative example to illustrate misalignments between the reference points and the registered landmarks is provided in Figure~\ref{fig:reg_scatter}.

Regarding point metrics, the ellipse (\textbf{E}) and ellipse with affine methods (\textbf{E+A}) obtain overall better results than the other methods. In general, the differences observed were found to be statistically significant for both point metrics and most anatomical structures. Exceptions to this are the thalami and the cerebellum where the pixel based affine registration method initialised using the reference ellipse (\textbf{AFF+I}), obtains results that appear worse but are not significantly different. Comparing the \textit{E} and \textit{E+A} methods, small (and not statistically significant) differences can be observed. Using a refinement registration after ellipse-based method slightly worsens the metrics when applied to the skull but improves them in all other anatomical structures. This is an expected result as the main focus of the first method is to co-register the skulls (ellipses). The uninitialised affine transformation (\textbf{AFF}) obtains significantly worse results than these two methods and has a higher variance of values as illustrated by its wide boxplots. Both methods using exclusively pixel-wise affine registration (\textit{AFF}, \textit{AFF+I}) achieve results that are worse than the metrics of the original moving image in some cases. This illustrates the disconnect between pixel-based metrics guiding these methods and point-based metrics targeting distances between real anatomical structures (especially for noisy sequences).

Regarding the differences between the Euclidean and Hausdorff metrics (especially the median values shown by the central line in boxplots), slightly higher Hausdorff values indicate that some point pairs are further than the average euclidean distance mean value. This is especially noticeable on the skull for \textit{E} and \textit{A+E} and the thalami for \textit{AFF}. 

The other two metrics show the same general tendencies, even though they focus on different aspects of the registration. The low SSIM values observed for all methods (even though \textit{E} and \textit{A+E} outperform all methods) illustrats how challenging this registration scenario is. The variations between individuals and acquisitions and low SNR make the intensity values particularly unreliable. On the other hand, the high anatomical DSC results observed for the ellipse-based methods (includding \textit{AFF+I}) validate our geometric approach that relies on a rough initial skull estimate to determine the general shape and orientation of the brain. Similarly high results obtained for the cerebellum indicate that a correct skull placement is a crucial important step towards the registration of all brain structures. 

\begin{figure}[ht]
\centering
\includegraphics[width=0.8\textwidth]{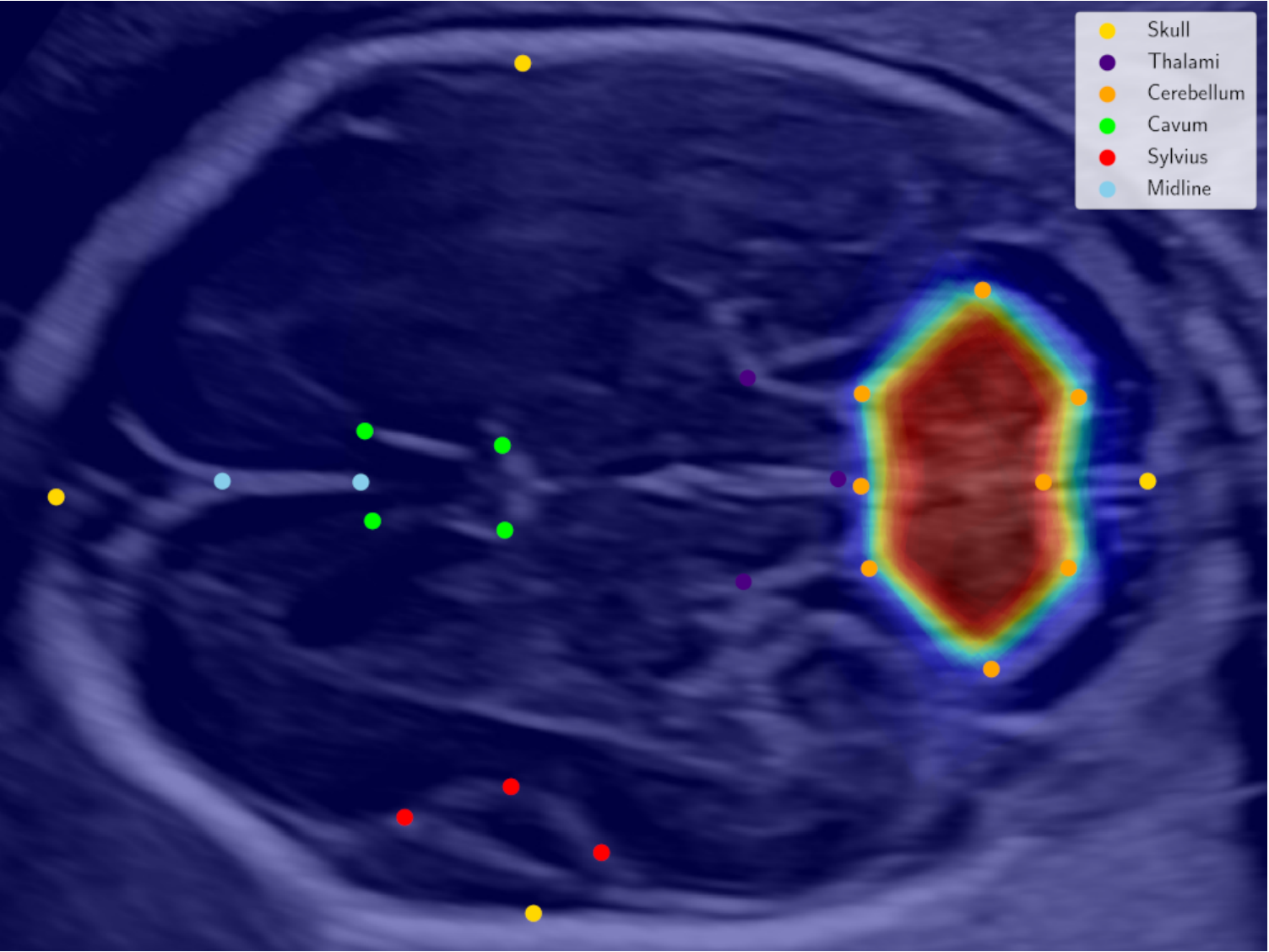}
\caption{Probabilistic map for the cerebellum structure based on the averaged concave hull of the landmarks for each image. Landmarks for the reference image (background) are also provided.}\label{fig:atlas}
\end{figure}

Finally, Figure~\ref{fig:atlas} shows the final heatmap generated from the registered concave hulls (\textit{E+A}) of the cerebellum for all output images. Even though faint outlines of some incorrect registration results outside of the cerebellum region (delimited by orange points) can be observed, the higher probability regions in the heatmap clearly correspond to the cerebellum region of the reference image delimited by the manually annotated orange landmarks.

\section*{Usage Notes}



The code provided to analyse the images and perform a rough initial alignment to a common space has been developed using python. A set of Jupyter notebooks detailing the use of each step is also provided in the repository with visualisation examples of each step of the processing pipeline. Furthermore, we provide the trained weights for the skull segmentation network as part of the data records (file \textit{unet.pt}~\cite{figshare_data}).

Regarding the use of python packages, the code heavily relies on pytorch (version 1.12.0 with CUDA 10.2) to do the heavy lifting. However, numpy (version 1.21.6), scipy (version 1.7.3), scikit-learn (version 1.0.2),  scikit-image (version 0.18.1), pandas (version 1.3.4), persim (version 0.3.1) and alphashape (version 1.3.1) are also used for some of the processing steps or to analyse different metrics related to the registration. Furthermore, opencv-python (version 4.5.2.52) and gimpformats (version 2021.1.4) where used to open the images with python. In particular, manual landmark annotation where done using the Gimp software and were stored (together with the ultrasound image) using the xcf format. Finally, matplotlib (version 3.7.1), seaborn (version 0.11.0) and statannot (version 0.2.3) were used for results visualisation inside the Jupyter notebooks.

\section*{Code availability}


All the code used in the study to generate the final atlas and to co-register the images is publicly available at \url{https://github.com/marianocabezas/fetal-brain}.




\section*{Author contributions statement}

All authors conceived and designed the experiments, C.M. and A.M collected the images and performed the landmark annotations, M.C. and Y.D. conducted the experiments and analysed the results. All authors reviewed the manuscript. 

\section*{Competing interests}

The authors declare no competing interests.

\section*{Figures \& Tables}

\end{document}